\documentclass[12pt]{amsart}

\theoremstyle{plain}

\newtheorem{theorem}{Theorem}[section]

\theoremstyle{definition}
\theoremstyle{definition}

\newcommand{\bean}{\begin{eqnarray}}
\newcommand{\eean}{\end{eqnarray}}
\newcommand{\ep}{\epsilon}

\newcommand{\ga}{\gamma}

\newcommand{\la}{\lambda}

\newcommand{\pa}{\partial}

\newcommand{\no}{\nonumber}

\begin{document}

 \title[Waterbag Model of dToda]
{Remarks on the waterbag model of dispersionless Toda Hierarchy}

\author[Jen-Hsu Chang]
{Jen-Hsu Chang \\
Department of Computer Science\\
 National Defense University\\
 Tauyuan, Taiwan\\
 E-mail: jhchang@ccit.edu.tw}
\maketitle

\begin{abstract}
We construct the free energy associated with the waterbag model of dToda. Also, the relations of
conserved densities are investigated. 
\end{abstract}
\section{Introduction}
The dispersionless Toda hierarchy (dToda) is defined by \cite{Ta1}
\begin{eqnarray}
\frac{\pa \mathcal{\lambda}}{\pa t_n} &=& \{B_n(p), \mathcal{\lambda} \}, \qquad
\frac{\pa \mathcal{\lambda}}{\pa \hat{t}_n}= \{ \hat{B}_n(p), \mathcal{\lambda} \},  \no \\
\frac{\pa \hat{\mathcal{\lambda}}}{\pa t_n} &=& \{B_n(p),\hat{\mathcal{\lambda}} \}, \qquad
\frac{\pa \hat{\mathcal{\lambda}}}{\pa \hat{t}_n}= \{ \hat{B}_n(p), \hat{\mathcal{\lambda}} \},
\quad n=1,2,3, \cdots \label{dto}
\end{eqnarray}
where the Lax operators $\lambda$ and $ \hat \lambda$ are
\begin{eqnarray*}
\mathcal{\lambda} &=& e^p+\sum_{n=0}^{\infty}u_{n+1}e^{-np} \\
\mathcal{\hat{\lambda}}^{-1} &=& \hat{u}_0 e^{-p}+ \sum_{n=0}^{\infty} \hat{u}_{n+1}e^{np}
\end{eqnarray*}
and
\[ B_n(p)=[\lambda^n]_{\geq 0}, \quad \hat B_n(p)=[\hat \lambda^{-n}]_{\leq -1}. \]
Here $[\cdots ]_{\geq 0}$ and $[\cdots ]_{\leq -1}$ denotes the non-negative part and negative
part  of  $\lambda^n$ and $\hat {\lambda}^{-n}$ respectively when expressed in the Laurent
series of $e^p$. For example,
\[ B_1(p)=e^p +u_1, \quad \hat {B}_1(p)= \hat{u}_0 e^{-p}.\]
Finally, the Poisson Bracket in \eqref{dto} is
\[ \{f(t_0,p), g(t_0,p)\}=\frac{\pa f}{\pa t_0}\frac{\pa g}{\pa p}-
\frac{\pa f}{\pa p}\frac{\pa g}{\pa t_0}. \]
\indent One can view $\lambda$ is a local coordinate near $"\infty"$ and $\hat \lambda$ as
a local coordinate near "0" \cite{kr2} \\
\indent According to dToda theory  \cite{Ta1}, there exist wave functions $S$, $\hat S$ and the dispersionless
$\tau$ function $F$ (or free energy)
\begin{eqnarray*}
S(\lambda)&=& \sum_{n=1}^{\infty}t_n \lambda^n+ t_0 \ln \lambda-\sum_{n=1}^{\infty}\frac{\pa_{t_n}F}
{n} \lambda^{-n}  \\
\hat S(\lambda)&=& \sum_{n=1}^{\infty} \hat{t}_n \hat{\lambda}^{-n}+ t_0 \ln \hat \lambda+\frac{\pa F}
{\pa t_0}- \sum_{n=1}^{\infty}\frac{\pa_{\hat t_n}F}
{n} \hat{\lambda}^{-n}
\end{eqnarray*}
such that
\begin{eqnarray*}
B_n(\lambda)&=& \pa_{t_n} S(\lambda) =  \lambda^n -\sum_{m=1}^{\infty}\frac{\pa^2_{t_n t_m}F}
{m} \lambda^{-m}  \\
B_n(\hat \lambda)&=& \pa_{t_n} \hat S(\hat \lambda)= \pa^2_{t_0 t_n}F - \sum_{m=1}^{\infty}
\frac{\pa^2_{t_n \hat t_m}F}{m} \hat \lambda^m \\
\hat{B_n} ( \lambda)&=& \pa_{\hat t_n}  S( \lambda)=  - \sum_{m=1}^{\infty}
\frac{\pa^2_{\hat t_n  t_m}F}{m}  \lambda^{-m} \\
\hat{B_n}(\hat \lambda) &=& \pa_{\hat t_n} \hat S(\hat \lambda)= \hat {\lambda}^{-n} +\pa^2_{t_0
\hat t_n}F - \sum_{m=1}^{\infty} \frac{\pa^2_{\hat t_n \hat t_m}F}{m} \hat \lambda^m. \\
\end{eqnarray*}
In particular,
\begin{eqnarray}
p(\lambda)&=& \pa_{t_0} S(\lambda) = \ln \lambda -\sum_{m=1}^{\infty}\frac{\pa^2_{t_0 t_m}F}
{m} \lambda^{-m} \no  \\
p(\hat \lambda)&=& \pa_{t_0} \hat S(\hat \lambda)= \ln \hat \lambda +\pa^2_{t_0 t_0}F
-\sum_{m=1}^{\infty}
\frac{\pa^2_{t_0 \hat t_m}F}{m} \hat \lambda^m. \label{pp}
\end{eqnarray}
Also,
\begin{eqnarray}
H_m^+ &=& \pa^2_{t_0 t_m}F=\frac{1}{m}\oint _{\infty} \lambda^m dp=
\frac{1}{m}\oint _{\infty} \lambda^m \frac{d \xi} \xi \no \\
H_m^- &=& \pa^2_{t_0 \hat t_m}F=\frac{1}{m}\oint _{0} \hat{\lambda}^{-m} dp=
\frac{1}{m}\oint _{0} \hat{\lambda}^{-m} \frac{d \xi} \xi , \quad m\geq1 \label{fr}
\end{eqnarray}
are the conserved densities of dToda hierarchy, where $p=\ln \xi $.
Then the dToda hierarchy \eqref{dto} can be expressed as
\begin{eqnarray}
\frac{\pa p(\lambda)}{\pa t_n}=\frac{\pa B_n(p(\lambda))}{\pa t_0}, \quad
\frac{\pa p(\lambda)}{\pa\hat  t_n}=\frac{\pa \hat{B_n}(p(\lambda))}{\pa t_0}, \no \\
\frac{\pa p(\hat \lambda)}{\pa t_n}=\frac{\pa B_n(p(\hat \lambda))}{\pa t_0}, \quad
\frac{\pa p(\hat \lambda)}{\pa \hat t_n}=\frac{\pa \hat{B_n}(p(\hat \lambda))}{\pa t_0},
\label{con}
\end{eqnarray}
$\lambda$, $\hat \lambda$ being fixed. The systems \eqref{con} are of  the conservational
laws for the dToda hierarchy. \\
\indent From \eqref{pp},one knows that
\[B_1(p)=e^p+u_1=e^p+\pa^2_{t_0 t_1} F, \quad \hat{B}_1(p)=\hat {u}_0e^{-p}=
e^{\pa^2_{t_0 t_0} F}e^{-p}.\]
Then from \eqref{con}, one has
\[p_{\hat t_1}=\pa_{t_0}[e^{\pa^2_{t_0 t_0} F}e^{-p}], \quad
 p_{ t_1}=\pa_{t_0}[e^p+\pa^2_{t_0 t_1} F].\]
Then $p_{t_1 \hat t_1}=p_{ \hat t_1 t_1}$ will imply
\[\pa^2_{t_1 \hat t_1}F=-e^{\pa^2_{t_0 t_0} F}.\]
It is the dToda equation. \\
\indent  This paper is organized as follows. In the next section, we construct the
waterbag model of dToda type from the Hirota equation.  Section 3 is devoted to finding
the free energy associated with the waterbag model from the Landau-Ginsburg formulation
in topological field theory. Also, the equations for conserverd densities
are obtained. In the final section, one discusses some problems to be investigated.
\section{Dispersionless Hirota Equation and Symmetry Constraints}
The dToda hierarchy \eqref{dto}(or \eqref{con}) is equivalent to the dispersionless Hirota
equation \cite{bm}:
\begin{eqnarray*}
D_{\mu} p(\lambda) &=&-\pa_{t_0} \ln [e^{p(\lambda)}-e^{p(\mu)}], \quad
\hat D_{\hat \mu} p(\lambda) =-\pa_{t_0} \ln [1-e^{p(\hat \mu)-p(\lambda)}] \\
D_{ \mu} p(\hat \lambda) &=&-\pa_{t_0} \ln [1-e^{p( \mu)-p(\hat \lambda)}], \quad
D_{\hat \mu} p(\hat \lambda) =-\pa_{t_0} \ln [e^{p(\hat \lambda)}-e^{p(\hat \mu)}] ,
\end{eqnarray*}
where
\[D_{\mu}=\sum_{m=1}^{\infty} \frac{\mu^{-m}}{m}\pa_{t_m},
\quad \hat D_{\hat \mu}=\sum_{m=1}^{\infty} \frac{\hat \mu^{m}}{m}\pa_{\hat t_m}. \]
We can also express them in terms of the $S$-function
\begin{eqnarray}
D_{\mu} S(\lambda) &=&- \ln [\frac{e^{p(\lambda)}-e^{p(\mu)}}{\mu}], \quad
\hat D_{\hat \mu} S(\lambda) =- \ln [1-e^{p(\hat \mu)-p(\lambda)}] \no \\
D_{ \mu} \hat S(\hat \lambda) &=&- \ln [1-e^{p( \mu)-p(\hat \lambda)}], \quad
\hat D_{\hat \mu} \hat S(\hat \lambda) =- \ln [\frac{e^{p(\hat \lambda)}-e^{p(\hat \mu)}}{\mu}].
\label{sf}
\end{eqnarray}
Next, we consider the symmetry constraints \cite{bk}.
\begin{itemize}
\item Case(I): $F_{t_0}=\sum_{i=1}^N \epsilon_i S_i,$ where $S_i=S(\lambda_i)$. \\
Then near "$\infty$" we have by \eqref{sf}
\begin{eqnarray*}
p &=& \ln \lambda-D_{\lambda} F_{t_0}=\ln \lambda-D_{\lambda}\sum_{i=1}^N \epsilon_i S_i \\
&=& \ln \lambda-\sum_{i=1}^N \epsilon_i D_{\lambda}S_i \\
&=& \ln \lambda-\sum_{i=1}^N \epsilon_i \ln [\frac{e^{p(\lambda)}-e^{h^i}}{\lambda}],
\quad h^i=p(\lambda_i) \\
&=& \ln \lambda-\sum_{i=1}^N \epsilon_i \ln [{e^{p}-e^{h^i}}]+ (\sum_{i=1}^N \epsilon_i)
\ln {\lambda}.
\end{eqnarray*}
Let $(\sum_{i=1}^N \epsilon_i)=0.$ Then one gets
\[\lambda=e^p \prod_{i=1}^N (e^p-e^{h^i})^{-\epsilon_i}.\]
Morever, near "0" we also have $\pa^2_{t_0 t_0}F=\sum_{i=1}^N \epsilon_i h^i$. Then
\begin{eqnarray*}
p(\hat \lambda) &=& \ln \hat \lambda+\pa^2_{t_0 t_0}F -\hat D_{\hat \lambda} F_{t_0} \\
&=& \ln \hat \lambda+\sum_{i=1}^N \epsilon_i h^i +\sum_{i=1}^N \epsilon_i \ln [1-e^{p(\hat \lambda)-{h^i}}] \\
&=& \ln \hat \lambda+\sum_{i=1}^N \epsilon_i h^i +\sum_{i=1}^N
\epsilon_i \ln [e^{-p(\hat \lambda)}-e^{-h^i}]+ (\sum_{i=1}^N \epsilon_i) p(\hat \lambda).\\
\end{eqnarray*}
Then one obtains
\[\hat \lambda=e^{p-\sum_{i=1}^N \epsilon_i h^i }\prod_{i=1}^N (e^{-p}-e^{-h^i})^{-\epsilon_i}.\]
Actually, we can see that
\begin{eqnarray}
\hat \lambda &=& e^{p} e^{-(\sum_{i=1}^N \epsilon_i)p} e^{-\sum_{i=1}^N \epsilon_i h^i }
\prod_{i=1}^N (e^{-p}-e^{-h^i})^{-\epsilon_i} \no \\
&=& e^p \prod_{i=1}^N (e^{h^i}-e^p)^{-\epsilon_i}= e^p \prod_{i=1}^N (e^p-e^{h^i})^{-\epsilon_i}
=\lambda . \label{eq}
\end{eqnarray}
Also,
\[H_1^+=\sum_{i=1}^N \epsilon_i e^{h^i}, \quad H_1^-=-e^{\sum_{i=1}^N \epsilon_i h^i}
(\sum_{i=1}^N \epsilon_i e^{-h^i}).\]
The evolutions for $t_1$ and $\hat t_1$ are
\begin{eqnarray}
\pa_{t_1} h^i &=& \pa_{t_0}[e^{h^i}+\sum_{i=1}^N \epsilon_i e^{h^i}] \no \\
\pa_{\hat t_1} h^i &=& \pa_{t_0}[e^{-h^i+\sum_{i=1}^N \epsilon_i h^i}] \label{wat}
\end{eqnarray}
We can also express them as the Hamiltonian form
\[
 \left[\begin{array}{c}
h^1 \\
h^2 \\
\vdots \\
 h^N
\end{array} \right]_{t_1}
=\eta^{ij} \pa_{t_0}
\left[\begin{array}{c}
\frac{\pa H_1^+}{\pa  h^1} \\
\frac{\pa H_1^+}{\delta  h^2}\\
\vdots \\
\frac{\pa H_1^+}{\delta h^N }
\end{array} \right], \quad
\left[\begin{array}{c}
h^1 \\
h^2 \\
\vdots \\
 h^N
\end{array} \right]_{\hat t_1}
=\eta^{ij} \pa_{t_0}
\left[\begin{array}{c}
\frac{\pa H_1^-}{\pa  h^1} \\
\frac{\pa H_1^-}{\delta  h^2}\\
\vdots \\
\frac{\pa H_1^-}{\delta h^N }
\end{array} \right], \]
where
\[ \eta^{ij}=\left[\begin{array}{ccccc}
1+\frac{1}{\ep_1} & 1 & \ldots & \ldots  & 1\\
1 &1+ \frac{1}{\ep_2} & 1 & \ldots & 1 \\
\vdots & \vdots & \ddots & \vdots & 1 \\
1 & 1 & \ldots & 1 &1+ \frac{1}{\ep_N}
\end{array} \right]. \]
\item Case(II): $F_{t_m}=\sum_{i=1}^N \epsilon_i S_i$. \\
Then
\begin{eqnarray*}
B_m(\lambda)&=&\lambda^m -D_{\lambda}F_{t_m}=\lambda^m -\sum_{i=1}^N \epsilon_i D_{\lambda} S_i \\
&=& \lambda^m -\sum_{i=1}^N \epsilon_i \ln (e^p-e^{h^i}).
\end{eqnarray*}
Hence the Lax operator is
\begin{eqnarray}
\lambda^m &=& e^{mp}+u_{m-1} e^{(m-1)p}+u_{m-2} e^{(m-2)p}+\cdots +u_{1} e^{p} \label{la2} \\
&+& u_0 +\sum_{i=1}^N \epsilon_i \ln (e^p-e^{h^i}). \no
\end{eqnarray}
\item Case (III): $ F_{\hat t_m}=\sum_{i=1}^N \epsilon_i S_i.$ \\
Then
\begin{eqnarray*}
\hat B_m(\hat \lambda)&=&\hat{\lambda}^{-m}+\pa^2_{t_0 \hat t_m}F -\hat D_{\hat \lambda}
F_{\hat t_m}= \hat \lambda^{-m}+\pa^2_{t_0 \hat t_m}F -\sum_{i=1}^N \epsilon_i \hat D_{\hat \lambda} S_i \\
&=& \hat \lambda^{-m} +\pa^2_{t_0 \hat t_m}F-\sum_{i=1}^N \epsilon_i \ln (e^{-p}-e^{-h^i}).
\end{eqnarray*}
Hence the Lax operator is
\begin{eqnarray}
\hat \lambda^{-m}&=&\hat u_m e^{-mp}+\hat u_{m-1} e^{-(m-1)p}+\hat u_{m-2} e^{-(m-2)p}+\cdots \label{la3} \\
 &+& \hat u_{1} e^{-p}+\hat u_0+\sum_{i=1}^N \epsilon_i \ln (e^{-p}-e^{-h^i}). \no
\end{eqnarray}
\end{itemize}

\section{Residue formula and free energy}
In this section, we compute the free energy associated with the waterbag model of case (I) in last
section, i.e., \eqref{eq}. Also, the relations of conserved densities are investigated. \\
\indent The free energy is a function $\mathbb{F}(t^1, t^2, \cdots, t^n)$ such that
 the associated functions,
\[c_{ijk}= \frac{\pa^3 \mathbb{F}}{ \pa t^i \pa t^j \pa t^k},\]
satisfy the following conditions.
\begin{itemize}
\item The matrix $\eta_{ij}=c_{1ij}$ is constant and non-degenerate.
 This together with the inverse matrix $\eta^{ij}$ are used to raise
 and lower indices.
\item The functions $c_{jk}^{i}=\eta^{ir}c_{rjk}$ define an associative
 commutative algebra with a unity element(Frobenius algebra).
\end{itemize}
\indent Equations of associativity give  a system of non-linear PDE
for $\mathbb{F}(t)$
\[\frac{\pa^3 \mathbb{F}(t)}{\pa t^{\alpha} \pa t^{\beta}
 \pa t^{\lambda}} \eta^{\lambda \mu} \frac{\pa^3 \mathbb{F}(t)}
 {\pa t^{\mu} \pa t^{\ga} \pa t^{\sigma}}  =
\frac{\pa^3 \mathbb{F}(t)}{\pa t^{\alpha} \pa t^{\ga}
 \pa t^{\lambda}} \eta^{\lambda \mu} \frac{\pa^3 \mathbb{F}(t)}
 {\pa t^{\mu} \pa t^{\beta} \pa t^{\sigma}}. \]
These equations constitute the Witten-Dijkgraaf-Verlinde-Verlinde (or WDVV)
equations. The geometrical setting in which to understand the free energy $\mathbb{F}(t)$
is the Frobenius manifold \cite{du1}. One way to construct such manifold is derived via
Landau-Ginzburg formalism as the structure on the parameter space $M$ of the appropriate
form
\[ \la=\la(p;t^1,t^2, \cdots, t^n).\]
The Frobenius structure is given by the flat metric
\begin{equation}
\eta( \pa, \pa')=-\sum res_{d \la =0} \{\frac{\pa(\la dp)\pa'(\la dp)}{ d \la (p)} \} \label{re1}
\end{equation}
and the tensor
\begin{equation}
c( \pa, \pa', \pa^{''})=-\sum res_{d \la =0} \{\frac{\pa(\la dp)\pa'(\la dp)\pa^{''}(\la dp)}
{ d \la (p) dp}  \} \label{re2}
\end{equation}
defines a totally symmetric $(3,0)$-tensor $c_{ijk}$. \\
\indent Geometrically, a solution of WDVV equation defines a multiplication
\[ \circ: TM \times TM \longrightarrow TM \]
of vector fields on the parameter space $M$, i.e,
\[ \pa_{t^{\alpha}} \circ \pa_{t^{\beta}}= c_{\alpha \beta}^{\gamma}(t) \pa_{t^{\gamma}}.\]
From $c_{\alpha \beta}^{\gamma}(t)$, one can construct intrgable hierarchies whose corresponding
Hamiltonian densities are defined recursively by the formula
\bean
\frac{\pa^2 \psi_{\alpha}^{(l)}}{\pa t^i \pa t^j}=c_{ij}^k \frac{\pa \psi_{\alpha}^{(l-1)}}
{\pa t^k}, \label{rec}
\eean
where $l \geq 1, \alpha =1,2, \cdots, n,$ and $ \psi_{\alpha}^{0}=\eta_{\alpha
\epsilon}t^{\ep}$. The integrability conditions for this systems are automatically
satisfied when the $c_{ij}^k$ are defined as above. \\
\indent In the following theorem,
{\bf one uses $\ln \lambda$ to replace  $ \lambda $}, which is the dual Frobenius
manifold associated with $\lambda$ \cite{du2, RS}.
\begin{theorem} Let the Lax operator be defined in \eqref{eq}.Then
\begin{eqnarray*}
&(I)& \eta(\pa_{h^i}, \pa_{h^j}) = \eta_{ij}=-\epsilon_i \epsilon_j, \quad i\neq j  \\
&(II)& \eta(\pa_{h^i}, \pa_{h^i}) = \eta_{ii}=-\epsilon_i^2+\epsilon_i  \\
&(III)& c(\pa_{h^i}, \pa_{h^j}, \pa_{h^k}) =c_{ijk}=\epsilon_i \epsilon_j \epsilon_k,\quad  i \neq j \neq k \\
&(IV)& c(\pa_{h^i}, \pa_{h^i}, \pa_{h^k}) = c_{iik}=\epsilon_i \epsilon_k [\epsilon_i +\frac
   {e^{h^k}}{e^{h^i}-e^{h^k}}] ,\quad  i \neq k \\
&(V)& c(\pa_{h^i}, \pa_{h^i}, \pa_{h^i}) = c_{iii}=\epsilon_i^3+ \epsilon_i [1-\epsilon_i
-\sum_{l=1, l\neq i}^N\frac{\epsilon_l e^{h^l}}{e^{h^i}-e^{h^l}}].
\end{eqnarray*}
\end{theorem}
\begin{proof}
We see that $\frac{\pa \ln \lambda}{ \pa h^i}=\epsilon_i \frac{e^{h^i}}{\xi-e^{h^i}}$, where
$p=\ln \xi$. Also, we have
\begin{equation}
\frac{d \ln \lambda}{d p}=1-\sum_{k=1}^N \epsilon_k \frac{e^{h^k}}{\xi-e^{h^k}}
=\frac{\prod_{k=1}^N (\xi-\omega_k)}{\prod_{k=1}^N (\xi-e^{h^k})}. \label{pol}
\end{equation}
In the following proofs, we use the  formula \eqref{pol} and the fact
that the residue at infinity is zero. \\
(I)
\begin{eqnarray*}
\eta(\pa_{h^i}, \pa_{h^j}) &=& \sum_{d \ln \lambda=0} Res \frac{\frac{\pa \ln \lambda}{ \pa h^i}
\frac{\pa \ln \lambda}{ \pa h^j}}{\xi \frac{d \ln \lambda}{d p}} d\xi  \\
&=& \sum_{d \ln \lambda=0} Res \frac{\epsilon_i \frac{e^{h^i}}{\xi-e^{h^i}}  \epsilon_j
 \frac{e^{h^j}}{\xi-e^{h^j}}}{\xi(1-\sum_{k=1}^N \epsilon_k \frac{e^{h^k}}{\xi-e^{h^k}})} d \xi \\
 &=&\sum_{d \ln \lambda=0} Res \frac{\epsilon_i \epsilon_j e^{h^i}e^{h^j}
  \prod_{k=1}^N (\xi-e^{h^k})}{\xi (\xi-e^{h^i})(\xi -e^{h^k}) \prod_{k=1}^N (\xi-\omega_k)} \\
  &=& -Res_{\xi =0}\frac{\epsilon_i \epsilon_j e^{h^i}e^{h^j}\prod_{k=1}^N (\xi-e^{h^k})}
  {\xi (\xi-e^{h^i})(\xi -e^{h^k}) \prod_{k=1}^N (\xi-\omega_k)}
= -\epsilon_i \epsilon_j ,\quad  i \neq j.
\end{eqnarray*}
(II) \begin{eqnarray*}
\eta(\pa_{h^i}, \pa_{h^i})&=&\sum_{d \ln \lambda=0} Res \frac{\epsilon_i^2 e^{2h^i}
  \prod_{k=1}^N (\xi-e^{h^k})}{\xi (\xi-e^{h^i})^2 \prod_{k=1}^N (\xi-\omega_k)} \\
  &=& -(Res_{\xi =0}+Res_{\xi=e^{h^i}})\sum_{d \ln \lambda=0} Res \frac{\epsilon_i^2 e^{2h^i}
  \prod_{k=1}^N (\xi-e^{h^k})}{\xi (\xi-e^{h^i})^2 \prod_{k=1}^N (\xi-\omega_k)} \\
  &=& -\epsilon_i^2 -\frac{\epsilon_i^2 e^{2h^i}\prod_{k=1,k\neq i}^N (e^{h^i}-e^{h^k})}
  {e^{h^i} \prod_{k=1}^N (e^{h^i}-\omega_k)}\\
  & =& -\epsilon_i^2-\epsilon_i^2 e^{h^i} \frac{1}{-\epsilon_i e^{h^i}}= -\epsilon_i^2+\epsilon_i
  \end{eqnarray*}
(III)\begin{eqnarray*}
 c(\pa_{h^i}, \pa_{h^j}, \pa_{h^k})&=& \sum_{d \ln \lambda=0} Res \frac{\epsilon_i e^{h^i}
   \epsilon_j e^{h^j} \epsilon_k e^{h^k} \prod_{l=1}^N (\xi-e^{h^l})}
   {\xi (\xi-e^{h^i})(\xi -e^{h^j})(\xi -e^{h^j}) \prod_{l=1}^N (\xi-\omega_l)} d \xi \\
   &=&- Res_{\xi=0} \frac{\epsilon_i e^{h^i}
   \epsilon_j e^{h^j} \epsilon_k e^{h^k} \prod_{l=1}^N (\xi-e^{h^l})}
   {\xi (\xi-e^{h^i})(\xi -e^{h^j})(\xi -e^{h^k}) \prod_{l=1}^N (\xi-\omega_l)} d \xi \\
   &=& \epsilon_i \epsilon_j \epsilon_k ,\quad i \neq j \neq k.
\end{eqnarray*}
(IV)
\begin{eqnarray*}
 c(\pa_{h^i}, \pa_{h^i}, \pa_{h^k})&=& \sum_{d \ln \lambda=0} Res \frac{\epsilon_i^2 e^{2h^i}
    \epsilon_k e^{h^k} \prod_{l=1}^N (\xi-e^{h^l})}
   {\xi (\xi-e^{h^i})^2(\xi -e^{h^k}) \prod_{l=1}^N (\xi-\omega_l)} d \xi \\
   &=&-[ Res_{\xi=0}+Res_{\xi=e^{h^i}}]\frac{\epsilon_i^2 e^{2h^i}
    \epsilon_k e^{h^k} \prod_{l=1}^N (\xi-e^{h^l})}
   {\xi (\xi-e^{h^i})^2(\xi -e^{h^k}) \prod_{l=1}^N (\xi-\omega_l)} d \xi \\
   &=&-[-\epsilon_i^2\epsilon_k +\frac{\epsilon_i^2 e^{2h^i}
    \epsilon_k e^{h^k} \prod_{l=1,l\neq i}^N (e^{h^i}-e^{h^l})}
   {e^{h^i}(e^{h^i} -e^{h^k}) \prod_{l=1}^N (e^{h^i}-\omega_l)} ] \\
   &=& -[-\epsilon_i^2 \epsilon_k -\frac{\epsilon_i \epsilon_k e^{h^k}}{e^{h^i}-e^{h^k}}]
   =\epsilon_i \epsilon_k [\epsilon_i+\frac{ e^{h^k}}{e^{h^i}-e^{h^k}}].
\end{eqnarray*}
(V)
\begin{eqnarray*}
 c(\pa_{h^i}, \pa_{h^i}, \pa_{h^i})&=& \sum_{d \ln \lambda=0} Res \frac{\epsilon_i^3 e^{3h^i}
  \prod_{l=1}^N (\xi-e^{h^l})}
   {\xi (\xi-e^{h^i})^3 \prod_{l=1}^N (\xi-\omega_l)} d \xi \\
   &=&-[ Res_{\xi=0}+Res_{\xi=e^{h^i}}]\frac{\epsilon_i^3 e^{3h^i}
  \prod_{l=1}^N (\xi-e^{h^l})}{\xi (\xi-e^{h^i})^3 \prod_{l=1}^N (\xi-\omega_l)} d \xi \\
   &=&-\{-\epsilon_i^3 +\epsilon_i^3 e^{3h^i}\frac{d}{d\xi}[ \frac{
   \prod_{l=1,l\neq i}^N (\xi-e^{h^l})}
   {\xi \prod_{l=1}^N (\xi-\omega_l)}] \mid_{\xi=e^{h^i}} \} \\
   &=& \epsilon_i^3+\epsilon_i^3e^{3h^i}\frac{1-\epsilon_i-\sum_{l=1, l\neq i}^N
   \frac{\epsilon_l e^{h^l}}{e^{h^i}-e^{h^l}}}{\epsilon_i^2e^{3h^i}} \\
   &=&\epsilon_i^3+\epsilon_i(1-\epsilon_i-\sum_{l=1, l\neq i}^N
   \frac{\epsilon_l e^{h^l}}{e^{h^i}-e^{h^l}}) .
\end{eqnarray*}
\end{proof}
Let's define $\Omega=\sum_{i=1}^N \frac{\pa}{\pa h^i}$. Then we can verify directly that
\begin{equation}
\eta(\pa_{h^i}, \pa_{h^j})=c(\pa_{h^i}, \pa_{h^j},\Omega)=
\sum_{k=1}^N c(\pa_{h^i}, \pa_{h^j},\pa_{h^k}). \label{uni}
\end{equation}
Also, from the Theorem, it's not difficult to check directly the compatibility
(or Egorov's condition)
\[\pa_{h^i} c_{lmn}=\pa_{h^l} c_{imn},\quad i,l,m,n=1 \cdots N.\]
Hence one can get the free energy associated with \eqref{eq}
\begin{eqnarray}
F(\vec {h})&=& \sum_{1\leq i < j < k \leq N} \epsilon_i \epsilon_j \epsilon_k h^i h^j h^k+ \frac{1}{6}
\sum_{i=1}^N (\epsilon_i-\epsilon_i^2+\epsilon_i^3)(h^i)^3 \no \\
&+& \frac{1}{2}\sum_{i  \neq k}^N \epsilon_i^2  \epsilon_k( h^i)^2 h^k \no \\
&+&\frac{1}{2}\sum_{ 1 \leq i < k \leq N} \epsilon_i \epsilon_k
[Li_3 (e^{h^i-h^k})+Li_3 (e^{h^k-h^i})], \label{free}
\end{eqnarray}
where $Li_3 (e^x)=\sum_{k=1}^{\infty} \frac{e^{kx}}{k^3}$ is the poly-logarithmic function.
Morever, from \eqref{uni}, one knows that $t^1=\sum_{i=1}^N h^i$.\\
\indent We remark that the free energy \eqref{free} is invariant under any permutation of
$(h^1, h^2, \cdots h^N)$. \\
\indent Furthermore, we have
\begin{eqnarray} c_{\alpha\beta}^{\gamma}&=& 0, \quad \alpha \neq \beta\neq \gamma;\quad
c_{\alpha\alpha}^{\beta}=\epsilon_{\alpha}\frac{e^{h^{\alpha}}}{e^{h^{\alpha}}-e^{h^{\beta}}},
\quad \alpha \neq \beta \no \\
c_{\alpha\beta}^{\beta}&=&\epsilon_{\beta}\frac{e^{h^{\beta}}}{e^{h^{\alpha}}-e^{h^{\beta}}},
\quad \alpha \neq \beta; \quad c_{\alpha\alpha}^{\alpha}=1-\sum_{\gamma\neq\alpha}
\epsilon_{\gamma}\frac{e^{h^{\gamma}}}{e^{h^{\alpha}}-e^{h^{\gamma}}}. \label{cij}
\end{eqnarray}
If we define $\phi_i=\frac{\pa \ln \lambda}{\pa h^i}=\epsilon_i \frac{e^{h^i}}{e^p-e^{h^i}}$
, then one has \begin{equation}
\phi_i \phi_j= c_{ij}^l \phi_l +Q_{ij}\frac{\pa \ln \lambda}{\pa h^i}, \label{ass}
\end{equation}
where \[Q_{ij}=\left\{
\begin{array}{ll}-\phi_i, & i=j,\\0, & i \neq j.\end{array} \right. \]
From \eqref{uni}, one knows that $\Omega$ is the unit element of the associative algebra
\eqref{ass}.\\
\indent Now, we have the following
\begin{theorem}Let $H_n^+$ and $H_n^-$ be the conserved densities defined in \eqref{fr}.
Then one has
\[
(I) \frac{\pa^2 H_n^+}{\pa h^i \pa h^j}=c_{ij}^k\frac{\pa H_n^+}{\pa h^k}, \quad
(II) \frac{\pa^2 H_n^-}{\pa h^i \pa h^j}=c_{ij}^k\frac{\pa H_n^-}{\pa h^k}. \]
\end{theorem}
\begin{proof}
(I)
\begin{eqnarray*}
\frac{\pa^2 H_n^+}{\pa h^i \pa h^j} &=& \frac{\pa}{\pa h^i}
\oint_{\infty} \la(\xi)^{n-1} \frac{\pa \la}{\pa h^j}\xi^{-1}d \xi
= \frac{\pa}{\pa h^i}
\oint_{\infty} \la(\xi)^{n-1} \lambda \frac{\pa \ln \lambda}{\pa h^j} \xi^{-1}d \xi  \\
&=&\frac{\pa}{\pa h^i}
\oint_{\infty} \la(\xi)^{n} \epsilon_j\frac{e^{h^j}}{e^p-e^{h^j}} \xi^{-1}d \xi  \\
&=& n
\oint_{\infty} \la(\xi)^{n}\epsilon_i\frac{e^{h^i}}{e^p-e^{h^i}}
 \epsilon_j\frac{e^{h^j}}{e^p-e^{h^j}} \xi^{-1}d \xi
+\oint_{\infty} \la(\xi)^{n}
 \epsilon_j\frac{\pa}{\pa h^i}(\frac{e^{h^j}}{e^p-e^{h^j}}) \xi^{-1}d \xi \\
&=& n
\oint_{\infty} \la(\xi)^{n}(c_{ij}^k\phi_k \xi^{-1}d \xi+Q_{ij} \frac{d \lambda}{\lambda})
-\oint_{\infty} \la(\xi)^{n}
 \epsilon_j\frac{\pa}{\pa p}(\frac{e^{h^j}}{e^p-e^{h^j}}) \xi^{-1}d \xi  \\
&=&n c_{ij}^k \oint_{\infty} \la(\xi)^{n}\phi_k \xi^{-1}d \xi
+n\oint_{\infty} \la(\xi)^{n-1}Q_{ij} d \lambda  \\
&-&\oint_{\infty} \la(\xi)^{n}
 [\frac{\pa}{\pa p}(\epsilon_j\frac{e^{h^j}}{e^p-e^{h^j}})] \xi^{-1}d \xi  \\
&=& c^k_{ij} \frac{\pa}{\pa h^k} \oint_{\infty} \la^{n} \xi^{-1} d\xi
-\oint_{\infty}\frac{\pa}{\pa p}[\la(\xi)^{n}\epsilon_j\frac{e^{h^j}}{e^p-e^{h^j}}]  dp \\
& =&c^k_{ij} \frac{\pa H_n^+}{\pa h^k} , \quad  n \geq  1.
\end{eqnarray*}
(II) The calculation is similar.
\end{proof}
\section{Concluding remarks}
We find the free energy associated with the waterbag model \eqref{eq} using the Landau-Ginsburg
formulation. From the free energy, one can establish the equations for the conserved densities
$H_n^+$ and $H_n^-$. When comparing the watebag model of dKP \cite{ch2}, one can't construct
the recursive operator of $H_n^+$ or $H_n^-$ from Theorem 3.2. Therefore, the bi-Hamiltonian
structure of \eqref{wat} is still unknown. On the other hand, we can construct the integrable
hierarchy via \eqref{rec} and $\eqref{cij}$; however, it won't be the dToda hierarchy. Finally,
finding the free energies  associated with \eqref{la2} and \eqref{la3} is also
very interesting. \\
{\bf Acknowledgments\/} \\
The author is grateful to Professors Maxim V.Pavlov and Jyh-Hao Lee for their useful discussions.
The work is supported by the National Science Council under grant no. NSC 96-2115-M-606-001-MY2.

\end{document}